\newcommand\useBBL{TT}
\newcommand\showchange{FL}
\newcommand\showremove{TT}
\documentclass[conference]{IEEEtran}
\IEEEoverridecommandlockouts

\usepackage{multirow}
\usepackage{cite}
\usepackage{textcomp}
\usepackage[misc]{ifsym}

\usepackage[braket]{qcircuit}
\usepackage{tikz}
\usetikzlibrary{calc}
\usetikzlibrary{backgrounds}
\usetikzlibrary{math}
\newcommand{\qn}[1]{{#1_1,\dots,#1_n}}
\newcommand{\seqn}[1]{{#1_1\dots #1_n}}

\usepackage{xcolor}

\usepackage{soul}
\newcommand{\bluet}[1]{{\color{blue} #1}} 
\newcommand{\redt}[1]{{\color{red} #1}} 
\if\showremove
\newcommand{\remv}[1]{{\color{magenta}\st{#1}}}
\else
\newcommand{\remv}[1]{{}} 
\fi
\if\showchange

\newcommand{\yu}[1]{{\color{blue} #1}} 
\newcommand{\com}[1]{{\color{magenta} #1}}

\newcommand{\yuR}[2]{{\remv{#1}\yu{#2}}}
\else

\newcommand{\yu}[1]{{#1}} 

\newcommand{\yuR}[2]{{#2}}
\newcommand{\com}[1]{{#1}}
\fi

\def\bibbrev#1#2{#1}			

\newcommand{\bibconf}[3][]{#1 \bibbrev{#2}{#3 (#2)}}

\newcommand{\blind}[1]{\hl{~\_\_~}}		

\newcommand\ie{{\em i.e.,}}		
\newcommand\eg{{\em e.g.~}}		
\newcommand{\kw}[1]{{\textit{#1}}}		
\newcommand{\kn}[1]{{\texttt{\small #1}}}		
\newcommand{\knc}[1]{{\texttt{\scriptsize #1}}}		

\newcommand\ZY{{\bf YuZhang}}
\newcommand\DHW{{\bf HaoweiDeng}}

\usepackage{algorithm}  
\usepackage{algorithmicx}  
\usepackage{algpseudocode}  
\usepackage{amsmath}
\usepackage{ntheorem}
\theorembodyfont{\normalfont}
\newtheorem{definition}{Definition}
\usepackage{amssymb}
\usepackage{booktabs}
\usepackage{listings}
\usepackage{mathptmx}
\usepackage{multirow}
\usepackage{multicol}
\usepackage{float}
\usepackage{subfigure}
\usepackage{graphicx}
\usepackage{xspace}

\newcommand\mysys{\kn{QLifeReducer}\xspace}
\newcommand{\emp}{\textbf}

\begin{document}
\lstset{escapeinside={<@}{@>}}
\lstset{numbers=left,numberblanklines=false}
\newcommand*\DNumber{\addtocounter{lstnumber}{-1}}
\title{
\Huge
Optimizing Quantum Programs against Decoherence\\ 
\vspace{0.3em}\LARGE Delaying Qubits into Quantum Superposition
}
\author{\IEEEauthorblockN{Yu Zhang\IEEEauthorrefmark{1}, Haowei Deng, 
Quanxi Li, Haoze Song and Leihai Nie}
\IEEEauthorblockA{
  School of Computer Science and Technology\\
  University of Science and Technology of China, Hefei, 230027,China\\
  Email: \IEEEauthorrefmark{1}yuzhang@ustc.edu.cn, \{jackdhw,crazylqx,shz666,nlh\}@mail.ustc.edu.cn}
}

\maketitle
\pagenumbering{arabic}
\begin{abstract}
Quantum  computing  technology  has  reached  a  second  renaissance in the last decade.
However,
in the NISQ era pointed out by John Preskill in 2018,
quantum noise and decoherence,
which affect the accuracy and execution effect of quantum programs,
cannot be ignored
and corrected by the near future NISQ computers.
In order to let users more easily write quantum programs,
the compiler and runtime system
should consider underlying quantum hardware features such as decoherence.
To address the challenges posed by decoherence,
in this paper,
we propose and prototype \mysys to minimize the qubit lifetime in the
 input OpenQASM program by delaying qubits into quantum superposition.
\mysys includes three core modules,
\ie the parser, parallelism analyzer and transformer.
It introduces the layered bundle format to express the quantum program, 
where a set of parallelizable quantum operations is packaged into a bundle.
We evaluate quantum programs before and after transformed by \mysys on both real IBM Q 5 Tenerife and the self-developed simulator.
The experimental results show that \mysys reduces the  error  rate  of  a  quantum  program  when  executed  on IBMQ 5 Tenerife by 11\%; 
and can reduce the longest qubit lifetime as well as average qubit lifetime by more than 20\% on most quantum workloads.
\end{abstract}

\section{Introduction}
\label{sec:intro}
Quantum computing technology has reached a second renaissance in recent a decade.
In May of 2016,
IBM has made a 5-qubit superconducting chip available in the cloud to general public~\cite{ibmq201605}.
The possibility of programming an actual quantum device has elicited much enthusiasm.
Simultaneously quantum languages~\cite{abhari2012scaffold, steiger2018projectq,svore2018qsharp}, compilers~\cite{abhari2014scaffcc,cross2018ibmqiskit,haner2018compiling}, quantum instruction set architectures (QISA)~\cite{svore2006qasm,cross2017OpenQASM,smith2016quil,fu2019eQASM} and microarchitecture~\cite{fu2018QuMA}
have been studied by the academic community.
They still need to be developed to form
a full software stack 
in order to 
accelerate the development of quantum software and hardware.

As pointed out by Prof. Preskill in 2018~\cite{preskill2018nisq}, 
Noisy Intermediate-Scale Quantum (NISQ) technology will be available in the near future. 
The NISQ quantum computer with 
50-100 qubits may be able to perform
tasks which surpass the capabilities of today's classical computers 
but quantum noise such as 
decoherence in an entangled system
will limit the size 
of quantum circuits that can be executed reliably.
Due to the high overhead 
of quantum error correction~\cite{gottesman2010introQEC},
NISQ devices will not make use of it in the near term.
Therefore, 
as quantum 
test beds get larger,
quantum programming should be lifted to higher levels of abstraction, 
while
the 
compiler or runtime 
system
should consider the constraints of quantum hardware.

In quantum computing,
information is stored in quantum bits -- \kw{qubits}
and computation is performed by applying quantum gates
and measurements to the quantum state of qubits.
Quantum states are intrinsically delicate~\cite{gottesman2010introQEC}: 
on the one hand, 
quantum gates may introduce small errors 
which will accumulate; 
on the other hand,
looking at one quantum state will collapse it, 
called the loss of quantum coherence or \kw{decoherence}~\cite{zurek1991decoherence}.
The \kw{coherence time} is defined as the time during which a quantum state holds its superposition~\cite{metodi2011quantum:qcca}.
And each physical qubit has limited coherence time,
for example, to date, 
quantum states in promising superconducting quantum circuits only reach coherence times of up to 100\textmu s~\cite{reagor2016coherence}. 
In order to make better use of the fragile physical qubits,
research on parallelizing quantum circuits has been studied~\cite{broadbent2009parallelizing}\cite{matteo2016parallelizing}.
But more research is needed to explore optimization on 
quantum programs to fit underlying quantum hardware features.

To address the challenges posed by decoherence,
we propose a new approach to minimize the lifetime of each qubit in the quantum program by program analysis and transformation,
called \mysys (Qubit Lifetime Reducer).
Here, the \kw{lifetime} of a qubit is defined as starting from its first operation 
to the operation making it decoherent 
or the last one.
Since OpenQASM~\cite{cross2017OpenQASM} is a more popular and newly updated quantum circuit language,
we prototype \mysys to transform 
OpenQASM programs.
As shown in Fig.~\ref{fig:ex:qasm},
\mysys can decompose the \kn{h} gate 
operating on an array of qubits 
\kn{a} at line 3 of (a) into
two separate \kn{h} operations 
on each qubit 
at lines 3 and 5 of (b),
thus 
the lifetime of qubit 
\kn{a[1]} 
will be reduced and start after the \kn{measure} 
at line 4.
The shortening of a qubit's lifetime  can reduce error accumulated on the qubit,
so as to improve the accuracy of the quantum program.
Furthermore, the execution time of a quantum program might be shortened 
due to the shortened qubit lifetime.
\begin{figure}[htbp]
\centering
\footnotesize
\begin{minipage}[t]{.24\textwidth}
\centering
\begin{lstlisting}
qreg a[2];qreg b[1];              
creg c[3];           
<@\textbf{h a;}@>              
measure a[0]->c[0];<@\DNumber@>     

cx a[1],b[0];            
measure a[1]->c[1]; 
measure b[0]->c[2];
\end{lstlisting}
{(a) Before}
\label{fig:ex:qasm:a}
\end{minipage}
\hfill
\begin{minipage}[t]{.24\textwidth}
\centering
\begin{lstlisting}
qreg a[2];qreg b[1];              
creg c[3];           
<@\textbf{h a[0];}@>           
measure a[0]->c[0];     
<@\textbf{h a[1];}@>  
cx a[1],b[0];            
measure a[1]->c[1]; 
measure b[0]->c[2];    
\end{lstlisting}
{(b) After}
\label{fig:ex:qasm:b}
\end{minipage}
\caption{OpenQASM program example: reducing the lifetime of a[1]}
\label{fig:ex:qasm}
\end{figure}

The main contributions of this paper are as follows:

(1) 
We propose a layered approach to analyze the lifetime of qubits,
where each sequence of quantum operations possibly executed in parallel are packaged into a \kw{bundle},
accordingly forming the \kw{layered bundle format} of the program.

(2)
We design a transformation method
to determine 
which qubits' operations can be shifted back 
according to the layered bundle format, 
and then adjust them to obtain the transformed code,
thereby reducing the lifetime of these qubits.

(3)
We prototype \mysys to cope with OpenQASM programs by applying the methods proposed above, and evaluate it on both a real IBM Q\footnote{http://research.ibm.com/ibm-q/} 5 Tenerife quantum computer and self-developed quantum simulator for evaluation.

The evaluation results show that 
\mysys reduces the error rate by 11\% of 
a quantum program 
when executed on IBM Q 5 Tenerife; 
and can reduce the longest qubit lifetime 
as well as average qubit lifetime 
by more than 20\% on most quantum workloads. 
It also reduces the execution time of some quantum programs.
In addition,
the layered information generated by \mysys
can also provide a basis for further parallelization of quantum circuits.

The rest of the paper is organized as follows.
Section \ref{sec:pre} introduces quantum computing basics and 
motivation examples. 
Section \ref{sec:design} describes the design of \mysys including
the parallelism analysis and the transformer.
Section \ref{sec:eval} describes the evaluation, and
Section \ref{sec:concl} concludes.

\section{Quantum Computation}
\label{sec:pre}
This section first introduces basic concepts of quantum computation and quantum computer system stack,
then describes quantum decoherence and motivation examples against it.
\subsection{Quantum Computing Basics}
\label{sec:pre:qc}
The quantum bit, or qubit, has a state just as a classical bit.
It may be in an arbitrary \kw{superposition} of its two basis states labeled $\ket{0}$ (or $\ket{g}$, ground state) and $\ket{1}$ (or $\ket{e}$, excited state):
\begin{equation}
\small
\ket{\psi}=\alpha\ket{0}+\beta\ket{1}=\alpha\left[
\begin{array}{c}
     0  \\
     1 
\end{array}
\right]+\beta\left[
\begin{array}{c}
     1  \\
     0 
\end{array}
\right]=\left[
\begin{array}{c}
     \alpha  \\
     \beta 
\end{array}
\right]
\label{eq:superposition}
\end{equation}
with complex amplitudes $\alpha,\beta$ 
satisfying $|\alpha|^2+|\beta|^2=1$.
The state of a general $n$-qubit system can be an arbitrary superposition over all $2^n$ computational basis states, \ie
\begin{equation}
\small
\sum\limits_{\qn{q}\in\{0,1\}^n}c_{\seqn{q}}\ket{\seqn{q}}=\sum\limits_{i=0}^{2^n-1}c_i\ket{i}
\end{equation}
where the basis state $\seqn{q}$ is a binary number of integer $i$.
Again, the complex amplitudes $c_i$ should satisfy
$\sum_i|c_i|^2=1$.

\begin{figure}[htbp]
\centering{
\scriptsize{
\begin{minipage}[c]{.22\textwidth}
\centering  

$$
\begin{array}{rcl}
\vspace{0.5em}
\tt{Hardmard} &
\begin{tikzpicture}[scale=1.000000,x=1pt,y=1pt]
\draw[color=black] (0.000000,0.000000) -- (24.000000,0.000000);
\begin{scope}
\draw[fill=white] (12.000000, -0.000000) +(-45.000000:8.485281pt and 8.485281pt) -- +(45.000000:8.485281pt and 8.485281pt) -- +(135.000000:8.485281pt and 8.485281pt) -- +(225.000000:8.485281pt and 8.485281pt) -- cycle;
\clip (12.000000, -0.000000) +(-45.000000:8.485281pt and 8.485281pt) -- +(45.000000:8.485281pt and 8.485281pt) -- +(135.000000:8.485281pt and 8.485281pt) -- +(225.000000:8.485281pt and 8.485281pt) -- cycle;
\draw (12.000000, -0.000000) node {$H$};
\end{scope}
\end{tikzpicture} &
           \dfrac{1}{\sqrt 2}
           \begin{bmatrix}
           1 & 1	\\
           1 & -1  \\
           \end{bmatrix}
\\
\vspace{0.5em}
\tt{Pauli-X} &
\begin{tikzpicture}[scale=1.000000,x=1pt,y=1pt]
\draw[color=black] (0.000000,0.000000) -- (24.000000,0.000000);
\begin{scope}
\draw[fill=white] (12.000000, -0.000000) +(-45.000000:8.485281pt and 8.485281pt) -- +(45.000000:8.485281pt and 8.485281pt) -- +(135.000000:8.485281pt and 8.485281pt) -- +(225.000000:8.485281pt and 8.485281pt) -- cycle;
\clip (12.000000, -0.000000) +(-45.000000:8.485281pt and 8.485281pt) -- +(45.000000:8.485281pt and 8.485281pt) -- +(135.000000:8.485281pt and 8.485281pt) -- +(225.000000:8.485281pt and 8.485281pt) -- cycle;
\draw (12.000000, -0.000000) node {$X$};
\end{scope}
\end{tikzpicture} &
    \begin{bmatrix}
    0 & 1	\\
    1 & 0
    \end{bmatrix}
\\
\vspace{0.5em}
\tt{Pauli-Y} &
\begin{tikzpicture}[scale=1.000000,x=1pt,y=1pt]
\draw[color=black] (0.000000,0.000000) -- (24.000000,0.000000);
\begin{scope}
\draw[fill=white] (12.000000, -0.000000) +(-45.000000:8.485281pt and 8.485281pt) -- +(45.000000:8.485281pt and 8.485281pt) -- +(135.000000:8.485281pt and 8.485281pt) -- +(225.000000:8.485281pt and 8.485281pt) -- cycle;
\clip (12.000000, -0.000000) +(-45.000000:8.485281pt and 8.485281pt) -- +(45.000000:8.485281pt and 8.485281pt) -- +(135.000000:8.485281pt and 8.485281pt) -- +(225.000000:8.485281pt and 8.485281pt) -- cycle;
\draw (12.000000, -0.000000) node {$Y$};
\end{scope}
\end{tikzpicture} &
    \begin{bmatrix}
    0 & -i	\\
    i & 0
    \end{bmatrix}
\\
\vspace{0.5em}
\tt{Pauli-Z} &
\begin{tikzpicture}[scale=1.000000,x=1pt,y=1pt]
\draw[color=black] (0.000000,0.000000) -- (24.000000,0.000000);
\begin{scope}
\draw[fill=white] (12.000000, -0.000000) +(-45.000000:8.485281pt and 8.485281pt) -- +(45.000000:8.485281pt and 8.485281pt) -- +(135.000000:8.485281pt and 8.485281pt) -- +(225.000000:8.485281pt and 8.485281pt) -- cycle;
\clip (12.000000, -0.000000) +(-45.000000:8.485281pt and 8.485281pt) -- +(45.000000:8.485281pt and 8.485281pt) -- +(135.000000:8.485281pt and 8.485281pt) -- +(225.000000:8.485281pt and 8.485281pt) -- cycle;
\draw (12.000000, -0.000000) node {$Z$};
\end{scope}
\end{tikzpicture} &
    \begin{bmatrix}
    1 & 0	\\
    0 & -1
    \end{bmatrix}
\\
\vspace{0.5em}
\tt{Phase}   &
\begin{tikzpicture}[scale=1.000000,x=1pt,y=1pt]
\draw[color=black] (0.000000,0.000000) -- (24.000000,0.000000);
\begin{scope}
\draw[fill=white] (12.000000, -0.000000) +(-45.000000:8.485281pt and 8.485281pt) -- +(45.000000:8.485281pt and 8.485281pt) -- +(135.000000:8.485281pt and 8.485281pt) -- +(225.000000:8.485281pt and 8.485281pt) -- cycle;
\clip (12.000000, -0.000000) +(-45.000000:8.485281pt and 8.485281pt) -- +(45.000000:8.485281pt and 8.485281pt) -- +(135.000000:8.485281pt and 8.485281pt) -- +(225.000000:8.485281pt and 8.485281pt) -- cycle;
\draw (12.000000, -0.000000) node {$S$};
\end{scope}
\end{tikzpicture} &
    \begin{bmatrix}
    1 & 0  	\\
    0 & i  	\\
    \end{bmatrix}
\\
\vspace{0.5em}
\pi/8 &
\begin{tikzpicture}[scale=1.000000,x=1pt,y=1pt]
\draw[color=black] (0.000000,0.000000) -- (24.000000,0.000000);
\begin{scope}
\draw[fill=white] (12.000000, -0.000000) +(-45.000000:8.485281pt and 8.485281pt) -- +(45.000000:8.485281pt and 8.485281pt) -- +(135.000000:8.485281pt and 8.485281pt) -- +(225.000000:8.485281pt and 8.485281pt) -- cycle;
\clip (12.000000, -0.000000) +(-45.000000:8.485281pt and 8.485281pt) -- +(45.000000:8.485281pt and 8.485281pt) -- +(135.000000:8.485281pt and 8.485281pt) -- +(225.000000:8.485281pt and 8.485281pt) -- cycle;
\draw (12.000000, -0.000000) node {$T$};
\end{scope}
\end{tikzpicture} &
    \begin{bmatrix}
    1 & 0  	\\
    0 & e^{i\pi/4}  	\\
    \end{bmatrix}
\end{array}
$$
\label{fig:C:qpic1}
\end{minipage}
\hfill
\begin{minipage}[c]{.22\textwidth}
\vspace{4em}
controlled-NOT 
\\
(also called controlled-X,CX)
\vspace{1em}
~~\\
\vspace{2em}
\providecommand{\ket}[1]{\left|#1\right\rangle}
\begin{tikzpicture}[scale=1.000000,x=1pt,y=1pt]
\filldraw[color=white] (0.000000, -7.500000) rectangle (18.000000, 22.500000);
\draw[color=black] (0.000000,15.000000) -- (18.000000,15.000000);
\draw[color=black] (0.000000,15.000000) node[left] {$\ket{A}$};
\draw[color=black] (0.000000,0.000000) -- (18.000000,0.000000);
\draw[color=black] (0.000000,0.000000) node[left] {$\ket{B}$};
\draw (9.000000,15.000000) -- (9.000000,0.000000);
\filldraw (9.000000, 15.000000) circle(1.500000pt);
\begin{scope}
\draw[fill=white] (9.000000, 0.000000) circle(3.000000pt);
\clip (9.000000, 0.000000) circle(3.000000pt);
\draw (6.000000, 0.000000) -- (12.000000, 0.000000);
\draw (9.000000, -3.000000) -- (9.000000, 3.000000);
\end{scope}
\draw[color=black] (18.000000,15.000000) node[right] {$\ket{A}$};
\draw[color=black] (18.000000,0.000000) node[right] {$\ket{B \oplus A}$};
\end{tikzpicture}  
\\
\vspace{1em}
${
U_{CN}=
\begin{bmatrix}
1 & 0 & 0 & 0 	\\
0 & 1 & 0 & 0 	\\
0 & 0 & 0 & 1 	\\
0 & 0 & 1 & 0 	\\
\end{bmatrix}
}$  \\
\vspace{1em}
~~\\
\centering
\end{minipage}
}}
\caption{Names, symbols and unitary matrices for the common gates.}
\label{fig:gates}

\end{figure}
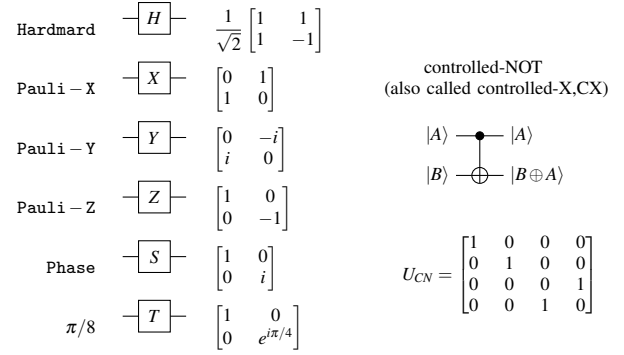

Quantum gates operate on qubits and change 
their state.
Fig.~\ref{fig:gates}
lists common one- and two-qubit
gates which are all reversible~\cite{nielsen2010quantum:qcqi};
that is,
each of them
can be described by a 
unitary matrix $U$,
where $U^{\dagger}U=1$ 
($U^\dagger$ is the adjoint of $U$).
An arbitrary $U$ operator on a 
qubit 
can be written as a combination of rotations, 
together with global phase shifts on the qubit~\cite{nielsen2010quantum:qcqi}:
\begin{equation}
\small
    U=e^{i\alpha}R_z(\beta)R_y(\gamma)R_z(\delta) 
\label{eq:ugate}
\end{equation}
The useful 2-qubit controlled-NOT (\kn{CNOT}) gate operates on 
a {control qubit} $\ket{c}$ and a {target qubit} $\ket{t}$),
performing 
$\ket{c}\ket{t}\rightarrow\ket{c}\ket{t\oplus c}$;
that is, if the control qubit is set to $\ket{1}$ then the
target qubit is flipped, otherwise the target qubit is left alone.
Multi-qubit gates are very hard to realize in hardware,
however,
they may be composed from \kn{CNOT} and single qubit gates~\cite{nielsen2010quantum:qcqi}.

Information stored in qubits is retrieved by measurements, which convert qubits into classical bits.
When measuring a qubit in the superposition state of Equation (\ref{eq:superposition}),
the outcome is either 0 or 1 with probability $|\alpha|^2$ or $|\beta|^2$, respectively, 
and 
the qubit collapses onto the basis state (\ket{0} or \ket{1}).

\subsection{Quantum Computer System Stack}
\label{sec:pre:qp}
\begin{figure}[htbp]
    \centering
    \includegraphics[width=0.35\textwidth]{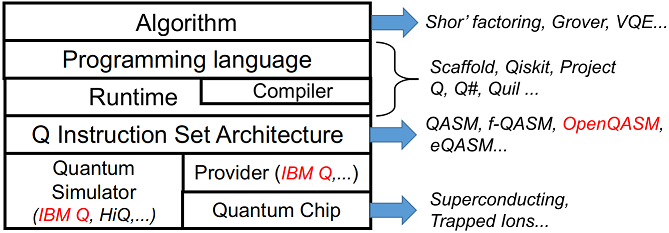}
    \caption{Overview of quantum computer system stack}
    \label{fig:qcstack}
\end{figure}
There has been a lot of research on each layer of the quantum computer system stack shown in Fig.~\ref{fig:qcstack}.
The recent introduced cloud access to quantum devices 
such as 
the IBM Q
~\cite{devitt2016ibmq}
~
and Qiskit\footnote{http://qiskit.org/}
~\cite{ibmq201703}
let users more easily write code
and run experiments on the provided
quantum devices and simulators
based on 
(Open)QASM quantum circuit languages~\cite{svore2006qasm,cross2017OpenQASM}.

The upper quantum algorithm 
can be described as a quantum-classical hybrid program containing a host program and multiple quantum kernels.
The host program can be written in a classic programming language such as C++ or Python,
and the quantum part is written in a high-level quantum programming languages such as Scaffold~\cite{abhari2012scaffold}
or
Project Q~\cite{steiger2018projectq}. 
The compiler infrastructure consists of a conventional host compiler such as GCC and a quantum compiler such as ScaffCC~\cite{abhari2014scaffcc}.
The quantum compiler works on the quantum part 
and generates quantum circuit IR
(intermediate representation)
belonging to a QISA. 
(Open)QASMs~\cite{svore2006qasm,cross2017OpenQASM,smith2016quil}
do not consider the low-level constraints 
to interface with the quantum processor.
They all lack control micro-architecture that implements and executes such instructions on a real quantum processor.
To bridge the gap between quantum software and hardware,
a quantum control micro-architecture 
QuMA~\cite{fu2018QuMA} and an executable QISA --
eQASM\cite{fu2019eQASM}
are proposed,
but  
only validated on a 2-qubit superconducting quantum processor.

\subsection{Quantum Noise and the Decoherence Problem} 
\label{sec:pre:decoherence}
Real quantum systems suffer from unwanted interactions with the outside world.
These unwanted interactions show up as \kw{noise} in quantum information processing (QIP) systems. 
For example,
both the entanglement of the quantum system with the surrounding environment and quantum measurements
will lead to the disappearance of quantum coherence,
denoted as \kw{quantum decoherence}.
Decoherence invalidates the quantum superposition principle and thus turns quantum computers
into (at best) classical computers, negating the
potential power offered by the quantumness of the
algorithms~\cite{zurek1991decoherence}.
To date, 
for the promising superconducting qubits, 
the longest coherence time is still within 10$\sim$100 \textmu s~\cite{reagor2016coherence};
a typical gate time is 20ns
for single-qubit gates 
and
$\sim$ 40 ns for 2-qubit gates, 
the duration of a measurement 
is typically 300ns - 1\textmu s~\cite{fu2019eQASM}.
Assume a single-qubit gate time is 
$\tau_u$,
a two-qubit gate time is 2$\tau_u$, 
and a measurement time is $\tau_m$ = $m\tau_u$, where $m$ is 15$\sim$50.

Due to decoherence,
the quantum program must complete execution quickly before the qubit state is decayed.
The longer a quantum program runs and the more operations it performs,
the more it is susceptible to noise.
Therefore,
it is necessary to shorten the duration of 
qubits in superposition in the program.

\subsection{Motivation Examples against Decoherence}
\label{sec:pre:ex}
We select OpenQASM to carry out the research in this paper 
since it is supported by IBM Q 
and can be generated by quantum compilers
such as ScaffCC.
Table~\ref{tab:openqasm} lists main quantum instructions in OpenQASM. 
The built-in universal gate basis is ``CNOT+$U(2)$''.
All 
the single qubit 
gates and 
two-qubit \kn{CNOT} gate shown in Fig.~\ref{fig:gates} are built in\footnote{
The definition of \kn{U} in OpenQASM is similar to Equation~(\ref{eq:ugate}), 
but without global phase shifts
on the qubit, 
\ie \kn{U} here is only $R_z(\beta)R_y(\gamma)R_z(\delta)$.
}.

\begin{table}[htp]
\centering
\caption{Main quantum-related statements in OpenQASM language}
\label{tab:openqasm}
\scriptsize
\begin{tabular}{ll}
\toprule
    \textbf{Statement} & Description\\
\midrule
    qreg name[size]; & Declare a named register of qubits\\
    gate name(params) qargs {body} &
    Declare a unitary gate\\
\midrule
    U($\gamma,\beta,\delta$) qubit$|$qreg; &
    Apply built-in single qubit gate(s) \\
    
    CX qubit$|$qreg,qubit$|$qreg; &
    Apply built-in CNOT gate(s)\\
    measure qubit$|$qreg -$>$ bit$|$creg; &
    Make measurement(s) in $Z$ basis\\
    gatename(params) qargs; &
    Apply a user-defined unitary gate\\
\bottomrule
\end{tabular}
\end{table}

\subsubsection{Lifetime of a qubit}

\begin{figure}[htbp]
\centering
\footnotesize
\begin{minipage}[t]{.21\textwidth}
\centering
\begin{lstlisting}
int i=0;
i=i+1;
<@\textcolor{blue}{int j=1;}@>
j+=i*2;
return j;
\end{lstlisting}
(a)
\label{fig:C:ex1}
\end{minipage}
\begin{minipage}[t]{.21\textwidth}
\centering
\begin{lstlisting}
int i=0,<@\textcolor{blue}{j=1}@>;
i=i+1;<@\DNumber@>

j+=i*2;
return j;
\end{lstlisting}
(b)
\label{fig:C:ex2}
\end{minipage}
\caption{C program examples: different declaration locations for variable \kn{j}}
\label{fig:ex:C}
\end{figure}

In the classical program such as C program
shown in Fig.~\ref{fig:ex:C},
moving \kn{int j=1} from line 3 of 
(a) to line 1 of (b) 
does not influence the execution result.
However,
as shown in Fig.~\ref{fig:ex:qasm} (a),
qubits \kn{a[0]} and \kn{a[1]}
are applied Hadamard operation at line 3,
entering superposition state,
and start their lifetime.
The lifetime of \kn{a[1]} ends after the measurement at line 6, 
occupying $(3+2m)\tau_u$.
During the period,
the \kn{measure} at line 4 is independent of \kn{a[1]},
however,
it makes the fragile qubit \kn{a[1]} 
have to wait on the superposition
before executing the \kn{CNOT} 
and accumulate error, 
accordingly increasing the program's error rate.
If decomposing \kn{h a;} 
and delaying \kn{h a[1];} at line 5 of Fig.~\ref{fig:ex:qasm} (b),
the lifetime of \kn{a[1]} will be reduced to $(3+m)\tau_u$.
Section~\ref{sec:eval} will show 
the movement improves the accuracy of the program 
running on a real IBM Q 5 Tenerife.

\subsubsection{Parallel Execution}
\label{sec:pre:par}
In order to better use the fragile physical qubits,
the parallelization of quantum circuits has been studied~\cite{broadbent2009parallelizing,matteo2016parallelizing}.
The recent eQASM~\cite{fu2019eQASM} 
adopts Single-Operation-Multiple-Qubit 
(SOMQ, similar to classical SIMD) execution, 
and a Very-Long-Instruction-Word (VLIW) architecture.
The former supports applying a single quantum operation on multiple qubits,
while the latter can combine multiple different quantum operations into a quantum bundle.
These parallel features need be considered 
when converting quantum programs into 
physical quantum circuits.
In this paper, we assume that neighboring gates operating on disjoint qubit subsets 
can always be applied in parallel, which is a common assumption for quantum technologies.

\begin{figure*}[htbp]
\centering
\scriptsize
\begin{minipage}[b]{.2\textwidth}
\centering
\begin{lstlisting}
qreg q[16];
creg c[16];
<@\textbf{h q[1];}@>
h q[2]; 
cx q[2], q[3];
h q[2]; 
<@\textbf{h q[6];}@>
cx q[6], q[11];
cx q[1], q[2];
h q[11]; 
h q[6]; 
cx q[6], q[11];
cx q[11], q[1];
...
\end{lstlisting}
{(a) Before}
\label{fig:ex:qasm:a}
\end{minipage}
\hfill
\begin{minipage}[b]{.23\textwidth}
\centering
\includegraphics[width=0.95\textwidth]{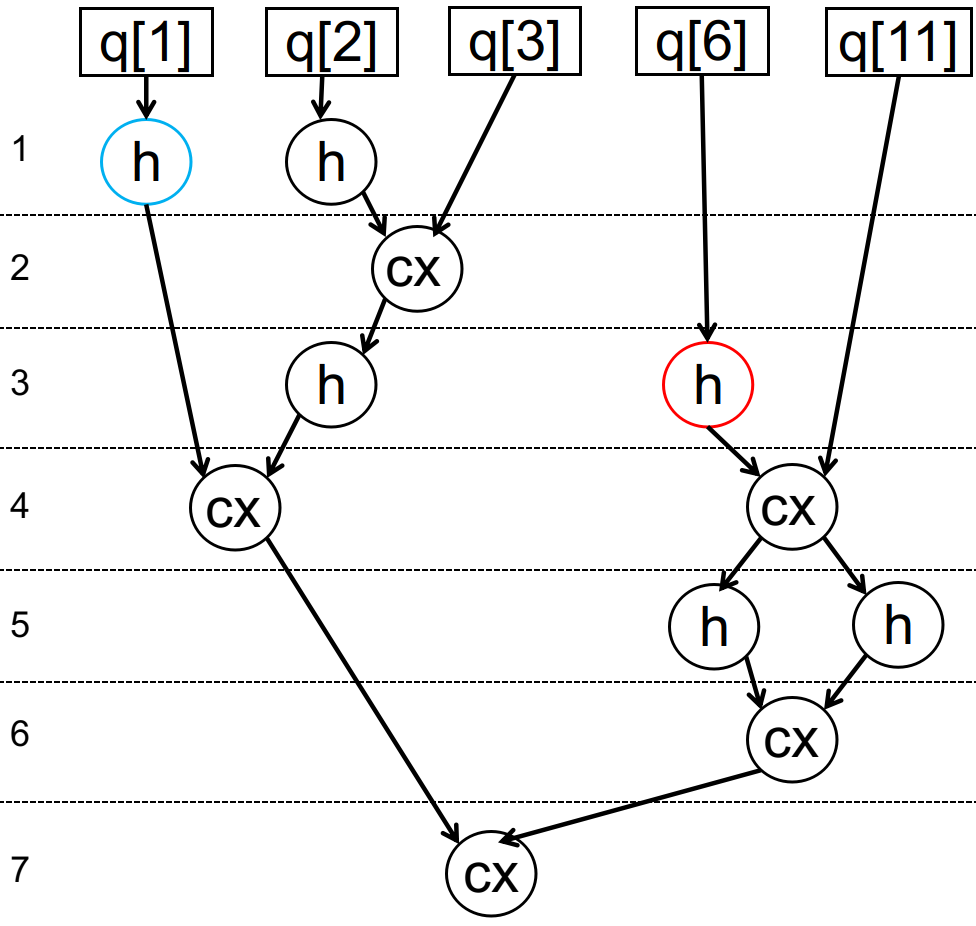}
\\\vspace{2em}
{(b) Before: Layered bundle format}
\label{fig:ex:qasm:b}
\end{minipage}
\hfill
\begin{minipage}[b]{.23\textwidth}
\centering
\includegraphics[width=0.95\textwidth]{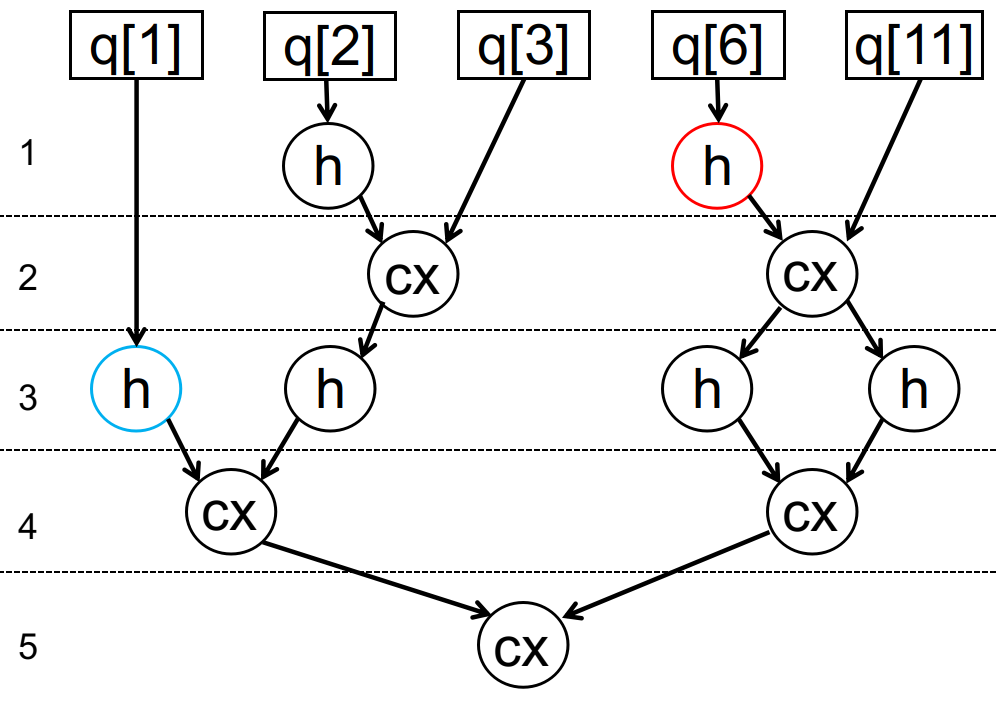}
\\\vspace{6em}
{(c) After: Layered bundle format}
\label{fig:ex:qasm:c}
\end{minipage}
\hfill
\begin{minipage}[b]{.2\textwidth}
\centering
\begin{lstlisting}
qreg q[16];
creg c[16];
h q[2];
<@\textbf{h q[6]; }@>
cx q[2], q[3];
cx q[6], q[11];
<@\textbf{h q[1]; }@>
h q[2]; 
h q[11]; 
h q[6]; 
cx q[1], q[2];
cx q[6], q[11];
cx q[11], q[1];
...
\end{lstlisting}
{(d) After}
\end{minipage}
\caption{OpenQASM program example: package operations into layered bundles}
\label{fig:ex:qasm3}
\end{figure*}

Take the quantum program in Fig.~\ref{fig:ex:qasm3} (a) as an example,
this code fragment is part of a workload provided by ProjectQ\cite{steiger2018projectq}, 
which is used to entangle a given number of 
qubits on IBM Q's 16-qubit quantum computer. 
Due to the parallelism of the quantum hardware and architecture,
we can analyze the dependencies of qubit operations in the program and package them into layered bundles,
where operations on the same layer is a bundle and can be executed in parallel.
Fig.~\ref{fig:ex:qasm3} (b)
shows the layered bundle format corresponding to code in Fig.~\ref{fig:ex:qasm3} (a).
The \kw{layered bundle format} (defined in Section \ref{sec:design:def}) reflects the parallelism and execution dependencies among quantum operations.
By further analysis, 
with the dependencies among quantum operations unchanged, 
the execution level of some quantum operations 
in Fig.~\ref{fig:ex:qasm3} (b)
can be adjusted to shorten the lifetime of the qubits involved,
accordingly obtaining Fig.~\ref{fig:ex:qasm3} (c). 
Thereinto,
node \bluet{\textcircled{h}} in column \kn{q[1]} (\kn{h q[1];}) is adjusted backward 
to layer 3,
node \redt{\textcircled{h}} in column \kn{q[6]} (\kn{h q[6];}) is adjusted forward 
to layer 1,
and the subsequent operations that depend on them are also adjusted.
The depth of the adjusted layered bundle format is reduced from 7 to 5. 
According to Section \ref{sec:pre:decoherence}, 
a layer that only contains single qubit gates such as \kn{h} costs $1\tau_u$, 
while the one contains two-qubit gates like \kn{cx} costs $2\tau_u$. 
So 
the total execution time of the program will be shortened from
$11\tau_u$ to $8\tau_u$
at maximum parallelization.
Similarly, we can calculate the lifetime of qubits. 
After the transformation, 
the lifetime of \kn{q[1]} reduces from $11\tau_u$ to $5\tau_u$.
By topologically sorting the nodes in Fig.~\ref{fig:ex:qasm3} (c), we can obtain the code sequence in Fig.~\ref{fig:ex:qasm3} (d). 

\section{Design}
\label{sec:design}
Following the idea of motivation examples described before, 
we have developed \mysys to reduce the qubit lifetime in a quantum program, 
considering the parallelism of quantum operations.
This section first gives an overview of the design 
then describes the related definitions 
and algorithms. 
\subsection{Overview}
\label{sec:design:overview}
\begin{figure}
    \centering    \includegraphics[width=0.48\textwidth]{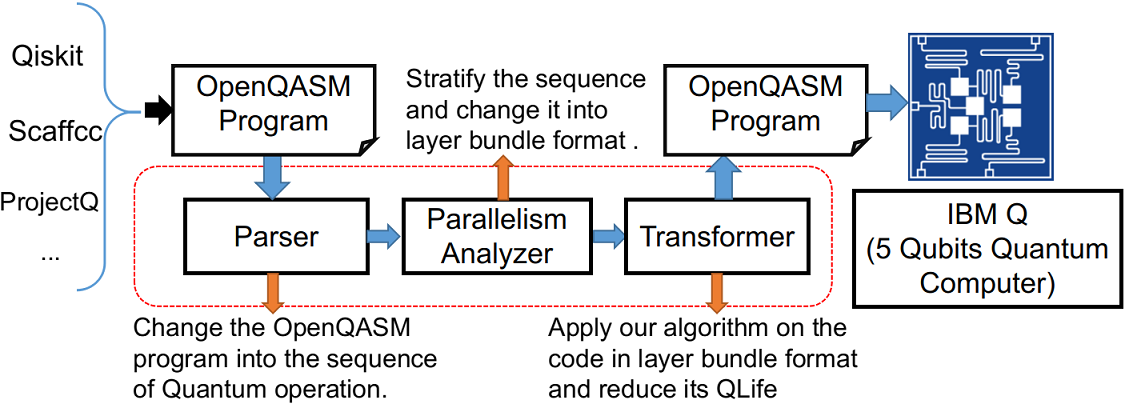}
    \caption{Overview of the \mysys}
    \label{fig:overview}
\end{figure}

The \mysys copes with an input OpenQASM program, 
and outputs the optimized OpenQASM code with shortened qubit lifetime.
The input program might come from ScaffCC\cite{abhari2014scaffcc}, 
ProjectQ\cite{steiger2018projectq}, QISKit\cite{cross2018ibmqiskit} or other OpenQASM provider.
As shown in Fig.~\ref{fig:overview}, 
there are three core modules in the \mysys.
The \kw{Parser} mainly performs macro expansion, 
that is, expanding user-defined gates (Table~\ref{tab:openqasm}), and obtains a sequence of built-in quantum instructions. 
The \kw{Parallelism Analyzer} (described in Section \ref{sec:design:parallel})
then analyzes the expanded sequence 
and packages each sequence of parallelizable operations into a bundle, forming the layered bundle format discussed in Section \ref{sec:pre:ex}. 
After that,
the \kw{Transformer} (described in Section \ref{sec:design:transformer}) analyzes the layered bundle format 
and adjusts it to reduce the qubit lifetime, 
then converts the adjusted one to the corresponding OpenQASM code.
The output OpenQASM program can be run on the quantum backend such as IBM Q quantum computer or simulator. 

In the following subsections,
we will use the example in Fig.~\ref{fig:ex:qasm3} to explain main algorithms of the \mysys.

\subsection{Definitions}
\label{sec:design:def}

\begin{definition}[
\textit{Qubit Set of an Instruction}]
     For a quantum instruction $\iota$,
     the \kw{qubit set} of $\iota$ is denoted as $S(\iota)$
\yuR{
     which is an ordered set 
}{}
     that contains the qubits operated by $\iota$. 
     For example, if $\iota$ is ``\kn{CX q[1],q[2]}'',
     then $S(\iota)$ = \{q[1], q[2]\}.
\yuR{
and \{q[1], q[2]\} is not equal to \{q[2], q[1]\}.
}{}
\end{definition}
\begin{definition}[\textit{Overlapped}]
     For two quantum instructions $\iota_1$ and $\iota_2$, if 
     \[
     S(\iota_1) \cap S(\iota_2) \neq \phi, 
     \]
     then they are \kw{overlapped} with each other.

     Overlapped instructions cannot be executed in parallel because only one quantum instruction can be applied on the
     intersecting qubit at the same time.
\end{definition}
\begin{definition}[\textit{Parallelizable}]
     For two quantum instructions $\iota_1$ and $\iota_2$, if 
     \[
     S(\iota_1) \cap S(\iota_2) = \phi,
     \]
     then they are \kw{parallelizable} with each other. 

     Parallelizable instructions can be executed in parallel.
\end{definition}
\begin{definition}[\textit{Bundle}]
     A \kw{bundle} is a set of quantum instructions that are parallelizable with each other. All instructions in one bundle can be executed in parallel. 

    The qubit set of a bundle is the union set of the qubit set of all the instructions in the bundle. 
\yu{
    For a bundle $b$,
     \[
     S(b) = \bigcup\limits_{\iota \in b} S(\iota) 
     \]
}
\end{definition}
\begin{definition}[\textit{Layered bundle format}]
    The \kw{layered bundle format} of a quantum program
\yu{
    can be represented as a directed acyclic graph,
    in which all the qubits involved are 
    start nodes at layer 0 and 
    instruction operators in each bundle are 
    nodes at the same layer like Fig.~\ref{fig:ex:qasm3} (b).
    Each directed edge $<o, \iota>$ connects 
    an instruction operator $\iota$ (arc head) and 
    another instruction operator or a qubit  $o$ (arc tail) 
    which $\iota$ directly depends on, 
    making instructions connected in execution order.\par
}
    We denote the layer of an instruction $\iota$ as $L(\iota)$.
    If there is directed edge $<n_2, n_1>$ in the layered bundle format,
    then node $n_1$ is the \kw{successor} of node $n_2$,
    if further satisfying $L(n_2)+1=L(n_1)$,
    then $n_1$ is the \kw{\yu{next-layer} successor} of $n_2$.
\end{definition}

\subsection{The Algorithm for Parallelism Analyzer}
\label{sec:design:parallel}
As shown in Fig.~\ref{fig:overview},
the Parallelism Analyzer is responsible for 
converting the quantum code 
(Fig.~\ref{fig:ex:qasm3}~(a)) 
into the 
layered bundle format (Fig.~\ref{fig:ex:qasm3}~(b)),
and the main algorithm is described 
in Alg.~\ref{alg:bundle}.
First, 
an empty array of bundles $B$ is initialized,
then it attempts to find a sequence of parallelizable instructions to form a bundle in each iteration of the outer {\bf while} loop.
During each iteration,
it first creates an empty bundle 
\yu{
$b$ as the current bundle
and an empty set $Q$ saving qubits operated 
by any instruction in $b$, 
then determines whether each instruction $I[index]$ processed in turn can be executed in parallel with instructions in $b$ 
by deciding whether the qubit intersection $Q\cap S(I[index])$ is empty.
If 
$I[index]$ can be  parallel with $b$, 
it will be added into $b$,
and the operated qubits will also be added into $Q$.
If the instruction is overlapped with 
$b$,
then $b$ is complete and can be added to $B$.
The iteration is then over and 
the next instruction will be handled in the next iteration.\par
} 
\begin{algorithm}
\footnotesize
 \caption{Transform code into layered bundle format}
 \label{alg:bundle}
 \begin{algorithmic}[2]
 \Require Array\ of\ quantum\ instructions: $I$
 \Ensure Array\ of\ bundles: $B$
 \Function {\it stratify}{Instruction\ I[]}
 \State $B \gets an\ empty\ array\ of\ bundles$
 \State $index \gets I.start$
 \While{$index \neq I.end$}
 \State $b \gets a\ new\ empty\ bundle$
 \State $Q \gets \phi$
 \While{$Q \cap S(I[index])=\phi$}
    \State $Q\gets Q\cup S(I[index])$
    \State $b\gets b \cup \{I[index]\}$
    \State $index \gets index+1$
 \EndWhile
 \State $B.append(b)$
 \EndWhile
 \State \Return{$B$}
 \EndFunction
 \end{algorithmic}
 \end{algorithm}
 
The code in layered bundle format will help us identify the instructions that might be shifted.
Furthermore,
it is also a basis of parallelizing quantum instructions 
and mapping them into parallel quantum circuits 
in our future work.

\subsection{Algorithms for the Transformer}
\label{sec:design:transformer}
\begin{figure}[htp]
\centering
  \includegraphics[width=0.48\textwidth]{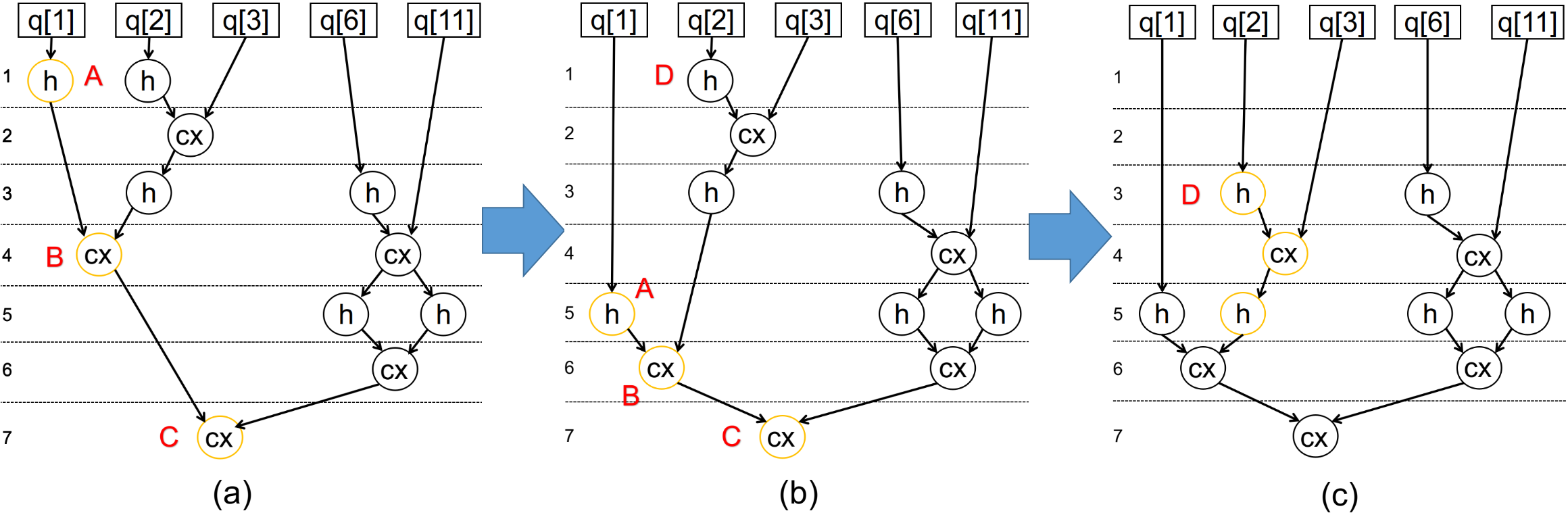}
\caption{Transformation process to reduce the qubit lifetime}
\label{fig:tranprocess}
\end{figure}

The Transformer performs two steps,
first adjusts the layered bundle format 
to shorten the qubit lifetime, 
\ie from Fig.~\ref{fig:ex:qasm3} (b) to (c);
then converts the adjusted layered bundle format 
to 
OpenQASM code, 
\ie from 
(c) to (d).
The first step is the core of the module and will be illustrated
via Fig.~\ref{fig:tranprocess} .

\subsubsection{Main Idea of the Adjustment}
First,
we need recognize instructions
that should be adjusted.
As discussed in Section \ref{sec:pre},
an instruction $\iota$ starts the lifetime
of a qubit $q$ only when it changes the qubit 
from ground state 
into superposition state. 
If the next instruction $\iota'$ operating on $q$
is not the \yuR{topological}{next-layer} successor of $\iota$,
$\iota$ could be delayed.
In Fig.~\ref{fig:tranprocess}~(a),
the \kn{h} gate operating on \kn{q[1]} at layer 1
(labelled \kn{A})
starts the lifetime of \kn{q[1]}.
Since \kn{A}'s \yuR{topological}{} 
successor (node \kn{B})
is at layer 4,
\kn{A} could be delayed to shorten the lifetime of \kn{q[1]}.

Second,
for an instruction $\iota$ to be delayed,
we need 
decide at which layer $\iota$ could be put,
and the adjustment must shorten the qubit lifetime 
without changing the execution order of 
\yuR{operations}{instruction operators} applied on a certain qubit.
We need iteratively analyze each successor $\iota'$ of $\iota$,
and ensure that the adjusted layer of $\iota$, 
\yu{denoted as $L'(\iota)$,}
is not less than the layer of other operand 
which $\iota'$ depends on.
Continue to consider Fig.~\ref{fig:tranprocess} (a),
\kn{B} (the successor of \kn{A}) has another operand \kn{q[2]} at layer 3, 
so $L'(\kn{A})\ge 3$ and $L'(\kn{A})\leq L'(\kn{B})-1$;
we further analyze \kn{B} and its successor \kn{C},
and obtain $L'(\kn{B})\ge 6$ and 
$L'(\kn{B})\leq L'(\kn{C})-1$ 
since \kn{C}'s another operand is \kn{q[6]} at layer 6.
Because \kn{C} has no successor, 
\kn{C} should not be adjusted, 
\ie $L'(\kn{C})=L(\kn{C})$.
By solving the above constraints,
we can get $L'(\kn{B})=6, L'(\kn{A})=5$ 
and the adjusted 
format is shown in Fig.~\ref{fig:tranprocess} (b).
Similarly,
we can adjust \kn{D} and get Fig.~\ref{fig:tranprocess} (c).

Third,
if the lowest layer that contains \yu{operators} is layer $n$ 
($n>1$),
then the layer and its successors 
could move forward to ($n-1$) layers.
For example,
layer 3 is the lowest layer 
that contains \yu{operators},
thus layers 3$\sim$7 could move forward to 1$\sim$5,
accordingly transforming 
from Fig.~\ref{fig:tranprocess} (c) 
to Fig.~\ref{fig:ex:qasm3} (c).

\subsubsection{Data Structures}
To implement the main idea,
we design data structures for qubits and instructions (or called operation).
A qubit is represented as a triple 
$q=(\yu{\kn{id}}, I, state)$, where
\begin{itemize}
    \item \yu{\kn{id} is the unique identifier of the qubit.}
    \item $I$ is an ordered list of instructions operating on the qubit. 
    The order of instructions in $I$ depends on their execution order in the original program.
    \item  $state$ is the qubit state,
\yu{
    whose value can be \kn{GROUND} or \kn{NOTGROUND},
    representing ground state and superposition state,
    respectively.
}
\end{itemize}

An instruction in the layered bundle format 
is represented as $\iota=(\kn{id}, op, seq, S, visited)$ contains four elements. 
\begin{itemize}
    \yu{\item \kn{id} is the unique identifier of the instruction.}
    \item $op$ is the operator of the instruction, \eg \kn{h}, \kn{cx}, \kn{measure}.
    \item $seq$ indicates the layer of the instruction,
    \eg \kn{A}$.seq=1$ in Fig.~\ref{fig:tranprocess} (a).
    \item $S$ is the qubit set of the instruction.
    \item $visited$ is a bool flag indicating whether the instruction has been visited.
\end{itemize}

\subsubsection{The Main Algorithm of the Transformer}

Alg. \ref{alg:trans} gives the definition of function {\it transform} that adjusts the code to reduce the qubit lifetime. 
It traverses all instructions in each bundle. 
For an unvisited instruction $\iota$, 
if \yu{any} of its operands is in \kn{GROUND} state, that is,
function $checkQ(\iota,\kn{NOTGROUND})$ returns false, 
instruction $\iota$ could be delayed. 
Then function {\it transform} will 
call function {\it adjust} to adjust related instructions. 
Alg. \ref{alg:setcheck} defines the auxiliary functions 
invoked by Alg. \ref{alg:trans} .

\begin{algorithm}
\footnotesize
 \caption{Transform code to reduce the qubit lifetime}
 \label{alg:trans}
 \begin{algorithmic}[2]
 \Require Array\ of\ bundles: $B$
 \Ensure Array\ of\ bundles\ with\ shorter\ qubit\ lifetime
 \Function {\it transform}{Bundle\ B[]}
 \State $index \gets B.start$
 \While{$index \neq B.end$}
 \ForAll{instruction\ $\iota \in B[index]$}
    \If{$(checkQ(\iota,\knc{NOTGROUND})=$ {True})} 
    \State continue
    \EndIf
    \If{$\iota.visited=$ {True}}
    \State continue    
    \EndIf
    \State $\iota.visited \gets$ {True}
    \State $setQ(\iota,\knc{NOTGROUND})$
    \State {\it adjust}$(\iota,B)$
 \EndFor
\State $index \gets index+1$
\EndWhile
\State $index \gets B.start$
 \com{\Comment{remove empty bundles at lower layers}}
\While{$B[index]\ is\ empty$}
 \State $index \gets index+1$
\EndWhile
\State $B.start\gets index$ \State \Return{$B$}
 \EndFunction
 \end{algorithmic}
 \end{algorithm}

\begin{algorithm}
\footnotesize
\setlength{\textfloatsep}{0.1cm}
\caption{Set and Check Qubit State}
\label{alg:setcheck}
\begin{algorithmic}
\Function{$setQ$}{Instruction $\iota$, State state}
\ForAll {$q \in S(\iota)$}
\State $q.state \gets state$
\EndFor
\EndFunction
\newline
\Function{$checkQ$}{Instruction $\iota$, State state}
\ForAll {$q \in S(\iota)$}
\If{$q.state \neq state$}
\State \Return{False}
\EndIf
\EndFor
\State \Return{True}
\EndFunction
\end{algorithmic}
\end{algorithm}
\setlength{\floatsep}{0.1cm}
\subsubsection{Algorithm on Adjustment}
\setlength{\textfloatsep}{0.1cm} 
\begin{algorithm}
\footnotesize
 \caption{Adjust an Instruction}
 \label{alg:adjust}
 \begin{algorithmic}[2]
 \Function{$adjust$}{Instruction $\iota$, Bundle $B[]$}
 \State $Q \gets S(\iota)$
 \State $cur \gets \iota.seq+1$ 
 \com{\Comment{$\iota\ belongs\ to\ B[\iota.seq]$}}
 \State $bstack \gets initialize\ an\ empty\ bundle\ stack$
 \State $bstack.push(\{\iota\})$
 \State $flag\gets$ False
 \While{$cur \neq B.end$ and $flag=$False}
 \State $b \gets a\ new\ empty\ bundle$
 \ForAll{$\iota_2 \in B[cur]$}
 \If{\yu{$\iota_2.visited=$False and} $S(\iota_2) \cap Q \neq \phi$}
 \If{$\iota_2.op=\knc{measure}$}
 \State $flag \gets$ True
 \State $setQ(\iota_2,\knc{GROUND})$
 \EndIf
 \State $b \gets b \cup \{\iota_2\}$
 \State $Q \gets Q \cup S(\iota_2)$ 
 \State $\iota_2.visited \gets$ True
 \State $B[cur] \gets B[cur] \setminus \{\iota_2\}$
 \com{\Comment{remove $\iota_2$ from $B[cur]$}}
 \EndIf
 \EndFor
 \If{$b\ is\ not\ empty$}
 \State $bstack.push(b)$
 \EndIf
 \State $cur \gets cur+1$
 \EndWhile
 \State $last \gets bstack.top()$
 \com{\Comment{The last instruction that enters the stack}}
 \State $line=last.seq$
 \While{$bstack\ is\ not\ empty$}
 \State $b \gets bstack.pop()$
 \State $B[line]\gets B[line] \cup b$
 \State $line \gets line-1$
 \EndWhile
 \EndFunction
 \end{algorithmic}
 \end{algorithm}

Alg.~\ref{alg:adjust} shows the pesudo code of function {\it adjust}
to do adjustment related to a given instruction $\iota$ 
in an array of bundles $B$. 
The adjustment should
keep the original execution order of operations related to each qubit,
and an instruction must not execute later than its successor. 
To adjust an instruction $\iota$, 
a bundle stack $bstack$ is introduced to save bundles of instructions to be adjusted,
and it is initialized as a stack with only one element $\{\iota\}$. 
A qubit set $Q$ is introduced to collect all qubits
depended by instructions saved in $bstack$,
and is initialized as $S(\iota)$.
Function {\it adjust} traverses and copes with all the successors of $\iota$ until reaching the end of the program 
or a \kn{measure} operation to any qubit in $Q$. 
When the function finds an instruction $\iota_2$ that is overlapped with $Q$, 
it means that $\iota_2$ is the successor of some instructions in $bstack$ and 
we need delay them together to keep the order constraints. 
So $\iota_2$ would be pushed into $bstack$ and 
$S(\iota_2)$ also be merged into $Q$. 
After the function finds all the instructions that need to be postponed, 
it will pop the stack and 
decide the new layer of each instruction according to the last one's layer. 

The $transform$ algorithm won't increase 
the number of bundles.
From Alg.~\ref{alg:adjust} you can see,
what we adjust on the quantum circuit program includes:
1) Remove an instruction $\iota_2$ from a bundle in the original quantum circuit,
and add $\iota_2$ into a temporary bundle $b$;
2) The formed temporary bundle is pushed into $bstack$,
and will be popped to adjust its layer later.

\subsection{Algorithm Complexity}
\yu{
Suppose a quantum circuit program has $n$ quantum instructions 
applied on $d$ qubits,  
$adjust(\iota, B)$
handles each unvisited instruction in $B[cur]$
( $cur>L(\iota)$ ),
thus the time complexity of Alg.~\ref{alg:adjust} is $O(n)$.
For function $transform(B)$,
it invokes $adjust()$ less than $d$ times, 
so the time complexity of Alg.~\ref{alg:trans} is $O(dn)$.
}

\section{Evaluation}
\label{sec:eval}

This section first introduces the \mysys prototype 
implementing algorithms mentioned in Section \ref{sec:design},
then evaluates the effect of \mysys 
on quantum program transformation, including accuracy and efficiency. 
\subsection{Prototype}
\label{sec:eval:prototype}
We have built \mysys and a simulator for evaluation on Linux with C++. 
As shown in Fig.~\ref{fig:overview}, 
\mysys takes the text of an OpenQASM program as input and outputs the transformed OpenQASM program.
First the Parser, 
where classes \kn{Instruction} and \kn{Qubit} are defined, 
analyzes the input text and builds the 
corresponding \kn{Instruction} list and \kn{Qubit} list. 
Then the Parallelism Analyzer,
where class \kn{Bundle} is defined,
analyzes the lists of \kn{Instruction} and \kn{Qubit}, 
and builds the array of \kn{Bundle}
according to Alg.~\ref{alg:bundle}. 
Finally, the Transformer 
analyzes the lists of \kn{Instruction} and \kn{Qubit} 
as well as the array of \kn{Bundle},
and outputs the transformed code 
according to Alg.~\ref{alg:trans} and \ref{alg:adjust}. 
The total number of LOC (lines of code) in \mysys is 1106. 

Since the availability of the real quantum computer is extremely limited, 
we build a simulator that calculates the lifetime of each qubit and the execution time about the input OpenQASM program
in the way discussed in Section~\ref{sec:pre:par}. 
The number of LOC in the simulator is 424.

\subsection{Methodology}
\label{sec:eval:method}
\subsubsection{Accuracy}
There is no available quantum simulator 
considering quantum noise
and the accuracy of programs when running on the simulators 
is always 100\%. 
So we use the real quantum device, IBM Q 5 Tenerife,
to evaluate the accuracy of two quantum circuits
before and after transformation in Fig.~\ref{fig:evalcircuit}. 
We discuss the detail in Section \ref{sec:eval:ibmq}.

\subsubsection{Qubit Lifetime}
The quantum hardware has some limitations.
It has serious error rate on quantum gates and measurements,
and only supports 5 qubits at most 
which is not enough for many quantum programs. 
Although the simulator cannot simulate quantum noise, 
it can calculate the execution time and the qubit lifetime 
of the program. 
Due to decoherence, 
these features would also affect the accuracy of quantum circuits. So we use our quantum simulator to evaluate 
the execution time, the longest qubit lifetime and average qubit lifetime of quantum programs.

\begin{table}[htp]
\centering
\caption{Quantum workloads}
\label{tab:workloads}
\footnotesize
\begin{tabular}{p{1.5cm}p{5.5cm}p{0.5cm}}
\toprule
    \bf Program & \bf Description &\bf Qubits \\
\midrule
    3G&
    3-qubit Grover's algorithm
    &  3\\
    \midrule
    DE &
    Deutsch's algorithm that exponentially accelerates classical algorithms
    & 2\\  
    \midrule
    4QFT, 5QFT& Quantum Fourier Transform using 4 or 5 qubits&4,5\\
    \midrule
    IBM 6 & Entangle 6 qubits in IBM's quantum chip and test their accuracy 
    & 16 \\    
    \midrule
    \kn{N}IQFT
    &Inverse Quantum Fourier Transform using \kn{N} qubits, where \kn{N} could be 4, 8, 16, 32, 64
    & \kn{N}\\
\bottomrule
\end{tabular}
\end{table}

Table~\ref{tab:workloads} lists
the tested quantum workloads, 
where {3G, DE, 4QFT and 5QFT come from Qiskit, IBM 6 from Project Q.  
Qiskit also provides 4IQFT, and we expand it with more qubits, 
\ie obtaining \kn{N}IQFT}.

\subsection{Experiments on Quantum Hardware}
\label{sec:eval:ibmq}
IBM Q has provided several superconducting quantum computers: 
IBM Q 20 Tokyo, IBM Q 14 Melbourne, 
IBM Q 5 Tenerife and IBM Q 5 Yorktown. 
Only Tenerife and Yorktown can be used for public,
but Yorktown is under maintenance. 
So we choose Tenerife for our experiments. 
Table~\ref{tab:tenerife} lists error rate and other parameters
for each quibit in Tenerife.
\begin{table}[htp]
    \footnotesize
    \caption{Parameters of IBM Q 5-qubit Tenerife}
    \label{tab:tenerife}
    \centering
    \begin{tabular}{p{3.4cm}|ccccc}
    \toprule
    & $Q_0$ & $Q_1$ & $Q_2$ & $Q_3$ & $Q_4$\\
    \midrule
    Frequency(GHz) & 5.25 & 5.30 & 5.35 & 5.43 & 5.18 \\
    \midrule
    Single-qubit gate error$(10^{-3})$ & 0.69 & 1.29 & 1.12 & 1.97 & 1.80 \\
    \midrule
    Readout error($10^{-2}$) & 6.10 & 6.90 & 7.90 & 7.80 & 25.20 \\
    \midrule
    Multi-qubit Gate error  & & CX10 & CX20 & CX32 &CX42 \\
    ($10^{-2}$)& & 3.22 & 2.59 & 7.46 & 5.53\\
    & & & CX21 & CX34 &\\
    &&& 4.23 & 6.73 &\\
    \bottomrule
    \end{tabular}
\end{table}

\begin{figure*}[htp]
\centering
\begin{minipage}[b]{.25\textwidth}
\centering
  \label{fig::before}
\footnotesize
  \includegraphics[width=\textwidth]{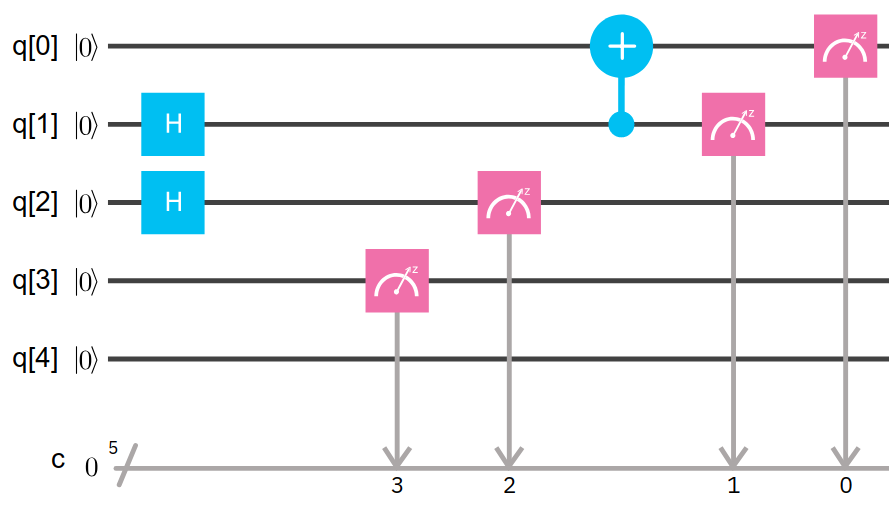}
{(a) Before transformation}
\end{minipage}
\hfill
\begin{minipage}[b]{.25\textwidth}
\centering
  \label{fig::after}
\footnotesize
  \includegraphics[width=\textwidth]{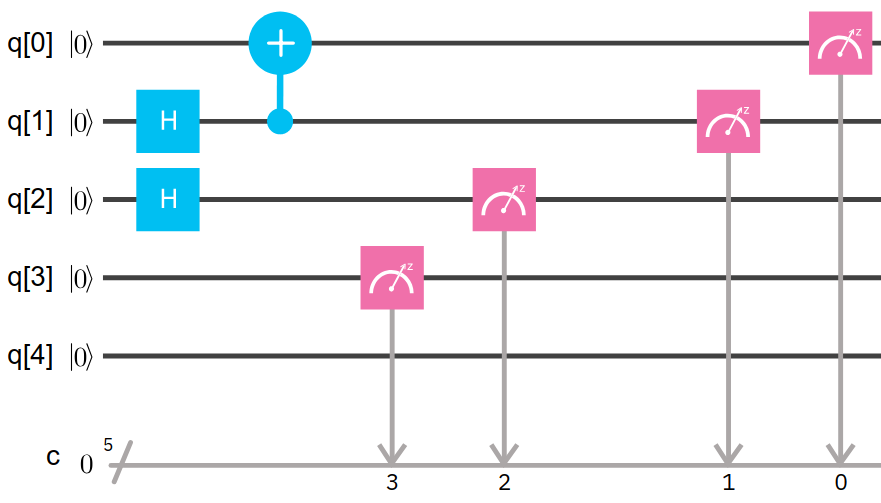}
{(b) After transformation}
\end{minipage}
\hfill
\begin{minipage}[b]{.26\textwidth}
\footnotesize
\begin{tabular}{p{1.6cm}|c|c}
    \toprule
        State & Before & After \\
    \midrule
       00 & 46.59\% & 46.01\%\\
       01 & 7.80\% & 7.52\%\\
       10 & 6.44\% & 5.12\% \\
       11 & 39.18\% & 41.35\%\\
    \midrule
       \textbf{Error 
       (01,10)} & 14.24\% & 12.64\% \\
    \bottomrule
    \end{tabular}
\\(c) Comparison between two circuits
\end{minipage}
\caption{Quantum circuits testing on IBM Q 5 Tenerife}
\label{fig:evalcircuit}
\end{figure*}

We use the accuracy of quantum circuits of Bell State\cite{nielsen2010quantum:qcqi} to show the benefits of \mysys. 
Fig.~\ref{fig:evalcircuit} (a) and (b) 
show the quantum circuits before and after transformation by \mysys,
and both of them will bound qubits \kn{q[0]} and \kn{q[1]} into Bell State. 
In the circuits, we focus on the result of qubits \kn{q[0]} and \kn{q[1]} and ignore the result of 
\kn{q[2]} and \kn{q[3]}. 
The only difference between the two circuits is 
the time gap between the \kn{H} gate on \kn{q[1]} and \kn{CNOT} 
on \kn{q[0]}, \kn{q[1]} which are used to create a Bell State. After 
the \kn{CNOT} operation, 
\kn{q[0]} and \kn{q[1]} are in Bell State and 
their state is $\frac{|00\rangle+|11\rangle}{\sqrt{2}}$. 
In an ideal quantum computer, 
the measured result of \kn{q[0]} and \kn{q[1]} should be 00 or 11 with both 50\% probability. 
However, because of hardware error caused by the decoherence 
and quantum noise, 
there will be an amount of result 01 and 10
which are considered as error states.
To reduce the effect of outliers,
we run each circuit 1024 times in one group on Tenerife,
and record the ratio of state $\ket{00},\ket{01},\ket{10}$ and $\ket{11}$.
25 groups for each circuit are tested on Tenerife,
and the average ratios of each resulting state 
are listed on Fig.~\ref{fig:evalcircuit} (c).
We see that the error rate (ratio of error states) 
reduces from 14.24\% 
down to 12.64\%. 
From the experimental results,
we see that IBM's quantum chips have significant noise,
where the final error rate is higher than 10\%.
\mysys indeed improves the accuracy of quantum circuits against decoherence for quantum computers with noise.

\subsection{Evaluation on the Simulator}
\label{sec:eval:sim}
We run workloads listed in Table~\ref{tab:workloads} 
on the self-developed simulator and 
Table~\ref{tab:sim:result} lists the result. 

\begin{table}[htp]
\centering
\footnotesize
\caption{Test result of workloads obtained from the simulator}
\label{tab:sim:result}
\begin{tabular}{c|cc|cc|cc}
\toprule
        \multirow{2}*{\bf Workload} & \multicolumn{2}{c|}{\bf Execution time} & \multicolumn{2}{c|}{\bf Longest lifetime} & \multicolumn{2}{c}{\bf{Average lifetime}} \\
         \cline{2-7}
    & before & after & before & after & before & after \\
    \midrule
    3G & 128 &128 & 128 & 128 & 128 & 127\\
    DE & 20 & 20 & \emp{20} & \emp{19} & 4.8 & 4.6\\
    4QFT & \emp{103} & \emp{98} & \emp{103} & \emp{92} & 91.8 & 83.4\\
    5QFT & 126 & 126 & 126 & 126 & 126 & 104.4\\
    IBM 6 & \emp{306} & \emp{246} & \emp{300} & \emp{246} & 211 & 162 \\
    4IQFT & 68 & 68 & \emp{68} & \emp{34} & 42 & 29 \\
    8IQFT & 150 & 150 & \emp{150} & \emp{38} & 80 & 32 \\
    16IQFT & 362 & 362 & \emp{362} & \emp{46} & 172 & 37\\
    32IQFT & 978 & 978 & \emp{978} & \emp{62} & 420 & 46 \\
    64IQFT & 2978 & 2978 & \emp{2978} & \emp{94} & 1172 & 62\\
    \bottomrule
    \end{tabular}
\newline
\raggedright
The unit of values in Columns $2\sim7$ is $\tau_u$,
\ie the executiontime of single qubit gate, generally 20ns.
\end{table}

We see that \mysys reduce the average qubit lifetime 
for every workload,
and sometimes can reduce the total execution time of the program. The reduction is tiny when qubits number is less than 4(3G and DE). But as the number of qubits used in the program increases(other algorithms), 
the reduction of average qubit lifetime 
brought by \mysys is prominent. 
The reason is that there is spatial locality in quantum program. Programmers tend to use several certain qubits in one part and 
\mysys can reassemble the program and 
remove the spatial locality. 
This will help increase the program's parallelism 
and also reduce the qubit lifetime. 
With the reduction of qubit lifetime, 
quantum system can further effectively control the use of qubits.

\section{Conclusion and Future Work}
In this paper, we proposed \mysys that can reduce 
the qubit lifetime of quantum programs 
considering the parallel execution of quantum circuits. 
With a parallelism analyzer, 
\mysys converts the quantum program into layered bundle format. 
Then \mysys applies transformation algorithm on the code in layered bundle format and 
reduces its qubit lifetime and execution time. 
By shortening the qubit lifetime of the quantum program, 
\mysys can reduce the error rate caused by decoherence of qubits which is unavoidable in the NISQ quantum computers.

Our future work will include the efficiency improvement
of the transformation algorithms and 
the combination with other optimization methods. 
We are also intend to build a more comprehensive quantum simulator for evaluation 
that considers the noise in quantum hardware. 
\label{sec:concl}

\section*{Acknowledgment}
{\footnotesize
This work was partly supported by the grants of Anhui Initiative in Quantum Information Technologies (No. AHY150100) and the National Natural Science Foundation of China (No. 61772487). 
The authors also would like to thank Prof. Jinshi Xu for discussing quantum computing problems.}

\if\useBBL

\else
\bibliography{quantum}
\bibliographystyle{IEEEtran}
\fi
\end{document}